\documentclass[conference]{IEEEtran}
\IEEEoverridecommandlockouts
\usepackage{cite}
\usepackage{amsmath,amssymb,amsfonts}
\usepackage{hyperref}
\usepackage{algorithmic}
\usepackage{graphicx}
\usepackage{textcomp}
\usepackage{xcolor}
\usepackage{float}
\def\BibTeX{{\rm B\kern-.05em{\sc i\kern-.025em b}\kern-.08em
    T\kern-.1667em\lower.7ex\hbox{E}\kern-.125emX}}
    
\begin{document}


\title{Beyond the Leaderboard: Multi-Dimensional Evaluation of Dense and Mixture-of-Experts Models for Automated Program Repair\\
}

\author{
\IEEEauthorblockN{Anvi Kalpesh Shah}
\IEEEauthorblockA{
\textit{National Institute of Technology, Calicut}\\
anvi\_b230027cs@nitc.ac.in}
\and
\IEEEauthorblockN{Umamaheswara Sharma B}
\IEEEauthorblockA{
\textit{National Institute of Technology, Calicut}\\
busharma@nitc.ac.in}
}

\maketitle




\maketitle

\begin{abstract}
Automated Program Repair (APR) systems based on language models are commonly evaluated by whether generated patches pass the associated test suite. However, functional correctness alone can overlook differences in maintainability, security, and computational cost that affect the practical use of generated patches in real-world software engineering. This paper presents a multi-dimensional evaluation framework for APR that combines functional correctness, maintainability, security, and generation efficiency into a Weighted Quality Index (QI) inspired by the ISO/IEC 25010 software product quality model. We evaluate four code-specialized language models: three dense Qwen2.5-Coder models at 3B, 7B, and 14B parameters, and the 16B-parameter DeepSeek-Coder-V2-Lite Mixture-of-Experts (MoE) model with 2.4B active parameters. All models are evaluated locally on identical hardware to control for infrastructure-related differences in efficiency, using 40 bugs from QuixBugs and 90 bugs from Defects4J. Our results show that model rankings vary across the weighting schemes, highlighting the sensitivity of APR evaluation to the relative importance assigned to different quality dimensions. The MoE model achieves functional correctness that has no statistically significant difference (McNemar’s exact test) from both the 7B and 14B dense models despite using 3–6 times fewer active parameters, while the three dense models show a statistically significant scaling trend among themselves. These results suggest that active parameter count can provide a more informative lens than total parameter count for sparse code models, and that single-metric evaluation of APR quality can hide this kind of trade-off.
\end{abstract}


\section{Introduction}
Automated Program Repair (APR) aims to automatically generate patches for software defects\cite{xia2023}, reducing the effort required to diagnose and fix bugs. A common way of reporting APR performance is as a single number: the fraction of bugs for which a generated patch passes the associated test suite. While this provides a useful and reproducible measure of functional correctness, it does not capture the broader quality of the resulting patch. A patch that passes its tests but is unnecessarily complex, introduces a statically detectable security weakness, or requires substantially more computation to generate may be a worse engineering artifact than one that is simpler, safer, and more efficient—even when both are equally “correct” according to the test suite.\par
At the same time, the rise of sparse Mixture-of-Experts (MoE) architectures raises a related evaluation question: what does it mean to compare a “small” and a “large” model? A MoE model may have a large total parameter count but activate only a fraction of its parameters per token, making total parameter count an imperfect proxy for inference cost. Similarly, comparing models evaluated on different infrastructure—for example, a locally hosted model versus an API-hosted model—can make efficiency differences difficult to attribute solely to the models themselves. These considerations suggest that both parameter usage and evaluation infrastructure should be accounted for when examining efficiency across model architectures.\par
This paper addresses both issues together. We construct a Weighted Quality Index (QI) inspired by the ISO/IEC 25010 systems and software quality model\cite{iso25010}, which combines functional correctness with three characteristics ISO/IEC 25010 identifies as part of product quality: maintainability, security, and performance efficiency. We evaluate every model entirely on local hardware with an identical inference stack, removing the infrastructure confound, and we report both total and active parameter counts for every model so that architecture-driven efficiency claims can be distinguished from simple scale.\par
We ask two research questions:\\
\textbf{RQ1 (the quality framework).} Does evaluating APR patches through a multi-dimensional, ISO/IEC 25010-inspired Quality Index reveal information that functional-correctness-only evaluation misses, and how sensitive is the resulting ranking to the relative weighting of the four dimensions?\\
\textbf{RQ2 (dense vs. sparse models).} Within this framework, does a sparse (MoE) code model achieve patch quality comparable to dense models while using substantially fewer active parameters, and how does functional correctness change as dense models scale in parameter count?\vspace{0.6em}

Our contributions are threefold: (1) a reusable, ISO/IEC 25010-inspired multi-dimensional Quality Index methodology for APR evaluation that combines functional correctness, maintainability, security, and generation efficiency under configurable weighting schemes; (2) a fully local, infrastructure-controlled experimental evaluation across QuixBugs\cite{lin2017quixbugs} and Defects4J\cite{just2014defects4j}, covering 130 bugs and four code-specialized language models spanning dense and sparse architectures; and (3) an empirical analysis addressing both research questions, including statistical significance testing of the correctness comparisons and analysis of ranking sensitivity across QI weighting schemes.

\section{Related Work}

\textbf{APR Benchmarks.} QuixBugs provides small, algorithmic programs with known defects and is available in both Python and Java. Defects4J provides real-world bugs mined from open-source Java projects, together with tests that reproduce the corresponding defects. In our evaluation, we use 90 Defects4J bugs selected from candidates whose source-level fixes modify one to three non-test lines, focusing on relatively localized repairs while keeping the prompted code manageable for local inference. QuixBugs has been used to evaluate both traditional repair tools\cite{ye2018quixbugs} and LLMs\cite{prenner2021codex}\par
\textbf{Software Quality Models.} ISO/IEC 25010:2023 defines a software product quality model comprising nine quality characteristics\cite{iso25010}, including functional suitability, performance efficiency, security, and maintainability. Our Quality Index draws on four of these characteristics, operationalizing them for the evaluation of generated APR patches. Rather than treating the quality model as a direct measurement standard, we use it as a basis for structuring multiple dimensions of patch quality within a configurable composite index.\par
\textbf{Dense and Sparse Code Models.} Qwen2.5-Coder\cite{hui2024qwen} is a family of dense, code-specialized models with instruction-tuned variants spanning multiple parameter scales. DeepSeek-Coder-V2\cite{deepseek2024coderv2} uses a Mixture-of-Experts architecture in which only a subset of parameters is active for each token; the Lite variant contains 16B total parameters with approximately 2.4B active parameters. Our study applies this dense-versus-sparse comparison specifically to APR and evaluates the resulting models across multiple dimensions of patch quality.

\section{Methodology}
\subsection{Models}
We evaluate four instruction-tuned, code-specialized lan- guage models, summarized in Table I. All four models are quantized to run within the 12 GB VRAM of an NVIDIA RTX 4080 Laptop GPU using Ollama \cite{ollama} with context length =8192, on a machine with 32 GB of system RAM, with no CPU offloading. All models are evaluated on the same hardware using the same inference stack, helping to control for infrastructure-related differences when comparing generation efficiency.

\subsection{Benchmarks and Bug Selection}
We evaluate 130 bugs from two benchmarks: the 40-bug Java subset of QuixBugs and 90 bugs from Defects4J. For Defects4J, we selected bugs whose fixes modify one to three non-test source lines. From 107 qualifying candidates, we initially selected 100 using a fixed random seed. Ten of these were subsequently excluded because their prompts could not be processed reliably, leaving 90 bugs for the final evaluation. The selected Defects4J bugs come from 16 of the 17 standard projects, with Chart excluded because its revisions could not be resolved in the Git mirror used for this study.
For Defects4J, models are given the relevant method rather than the entire source file to keep the input manageable. The generated method is then inserted back into the original file before compilation and testing. This reduces the median prompt size by approximately 29 times.
Functional correctness on Defects4J is evaluated using the benchmark's trigger tests rather than the full project test suite. The implications of this choice are discussed in Section VI.

\begin{table}[H]
\caption{MODELS EVALUATED}
\label{tab:models}
\centering
\begin{tabular}{lccc}
\hline
Model & Total & Active & Arch. \\
\hline
Qwen2.5-Coder-3B-Instruct & 3B & 3B & Dense \\
Qwen2.5-Coder-7B-Instruct & 7B & 7B & Dense \\
Qwen2.5-Coder-14B-Instruct & 14B & 14B & Dense \\
DeepSeek-Coder-V2-Lite-Instruct & 16B & 2.4B & MoE \\
\hline
\end{tabular}
\end{table}

\subsection{Quality Dimensions}

Each patch is evaluated along four dimensions, with all resulting scores
normalized to the range $[0,1]$.

\textbf{Correctness.}
A patch receives a score of 1 if it compiles and passes the relevant tests,
and 0 otherwise. For QuixBugs, we use the complete test suite for each bug;
for Defects4J, we use the corresponding trigger tests. We use greedy decoding
with temperature 0 and generate one patch per bug (pass@1\cite{chen2021codex}) to improve reproducibility.

\textbf{Maintainability.}
We measure cyclomatic complexity using Lizard\cite{lizard} and compare the patched file
with the official reference fix. The maintainability score is

\[
M =
\exp\left(
-0.15\max\left(0, CCN_{\mathrm{patch}}-CCN_{\mathrm{ref}}\right)
\right),
\]

giving a score of 1 when the patch is no more complex than the reference and
decreasing as its complexity exceeds the reference. This dimension is
evaluated for patches that compile; non-compiling patches receive a score of 0.

\textbf{Security.}
We use SpotBugs 4.10.4\cite{spotbugs} with the FindSecBugs 1.14.0\cite{findsecbugs} plugin to identify
security-related static-analysis findings in the compiled patch and reference
fix. Only findings in the SECURITY category are considered. The security
score uses the same reference-relative decay function as maintainability:

\[
S =
\exp\left(
-0.15\max\left(0, F_{\mathrm{patch}}-F_{\mathrm{ref}}\right)
\right),
\]

where $F_{\mathrm{patch}}$ and $F_{\mathrm{ref}}$ denote the number of
security findings in the patched and reference versions, respectively.
Non-compiling patches receive a score of 0.

\textbf{Efficiency.}
We measure generation speed in tokens per second using the inference
engine's internal timing, excluding one-time model-loading latency. For each
model, efficiency is calculated as its mean generation speed across patches
and normalized relative to the fastest model:

\[
E(m) =
\frac{\operatorname{tok/s}(m)}
{\displaystyle\max_{m'} \operatorname{tok/s}(m')}.
\]

Maintainability and security are scored only for patches that successfully compile, regardless of whether they pass the correctness test; patches that do not compile score 0 on both dimensions.

\subsection{Quality Index Construction}

The QI for model $m$ is computed as the weighted sum of its scores across
the four quality dimensions:

\[
QI(m) = \sum_d w_d s_d(m),
\]

where $s_d(m)$ is the score of model $m$ on dimension $d$, and the weights
satisfy $\sum_d w_d = 1$. We report QI under three weighting schemes:
\textit{equal} ($w_d = 0.25$ for all dimensions),
\textit{correctness-priority} ($w_{\mathrm{correct}} = 0.5$, with the
remaining 0.5 divided equally among the other three dimensions), and
\textit{other-priority} ($w_{\mathrm{correct}} = 0.2$, with the remaining
0.8 divided equally). Reporting multiple schemes serves as a robustness
check rather than an attempt to identify a single ``true'' weighting.

We also observed an important methodological issue while constructing the
index. An earlier version of the efficiency normalization used min-max
scaling,

\[
E_{\mathrm{minmax}}(m) =
\frac{\operatorname{tok/s}(m)-v_{\min}}
{v_{\max}-v_{\min}},
\]

which assigns exactly 0 to the slowest model and 1 to the fastest regardless
of the magnitude of the underlying difference. Because the other three
dimensions were not renormalized in this way, this caused the efficiency
dimension to mechanically dominate the equal-weighted QI: the model with
the lowest correctness of the four briefly ranked first overall purely
because min-max normalization gave efficiency a full [0,1] range that no other dimension had. We therefore report results using the
ratio-to-fastest normalization described above, which preserves each
model's relative speed without forcing the slowest model to zero. This highlights the importance of normalization when combining different metrics.

\section{Results}

Table~II reports the four dimension scores for each model, aggregated
over all 130 bugs. Correctness ranges from 0.154 for Qwen2.5-Coder-3B
to 0.400 for Qwen2.5-Coder-14B. Maintainability and security are
comparatively close across the four models, ranging from 0.856--0.897
and 0.878--0.923, respectively. Efficiency ranges from 0.240 for
Qwen2.5-Coder-14B to 1.000 for Qwen2.5-Coder-3B, the fastest model
under the ratio-to-fastest normalization. Fig.~1 visualizes these
differences across the four dimensions.

\begin{table}[H]
\caption{PER-MODEL SCORES ACROSS ALL FOUR QUALITY DIMENSIONS ($n=130$ BUGS)}
\label{tab:dimension_scores}
\centering
\begin{tabular}{lcccc}
\hline
Model & Correct. & Maint. & Security & Effic. \\
\hline
Qwen2.5-Coder-3B & 0.154 & 0.856 & 0.878 & 1.000 \\
Qwen2.5-Coder-7B & 0.308 & 0.897 & 0.923 & 0.480 \\
Qwen2.5-Coder-14B & 0.400 & 0.880 & 0.900 & 0.240 \\
DeepSeek-Coder-V2-Lite & 0.323 & 0.877 & 0.900 & 0.810 \\
\hline
\end{tabular}
\end{table}

\begin{figure}[H]
\centering
\includegraphics[width=\columnwidth]{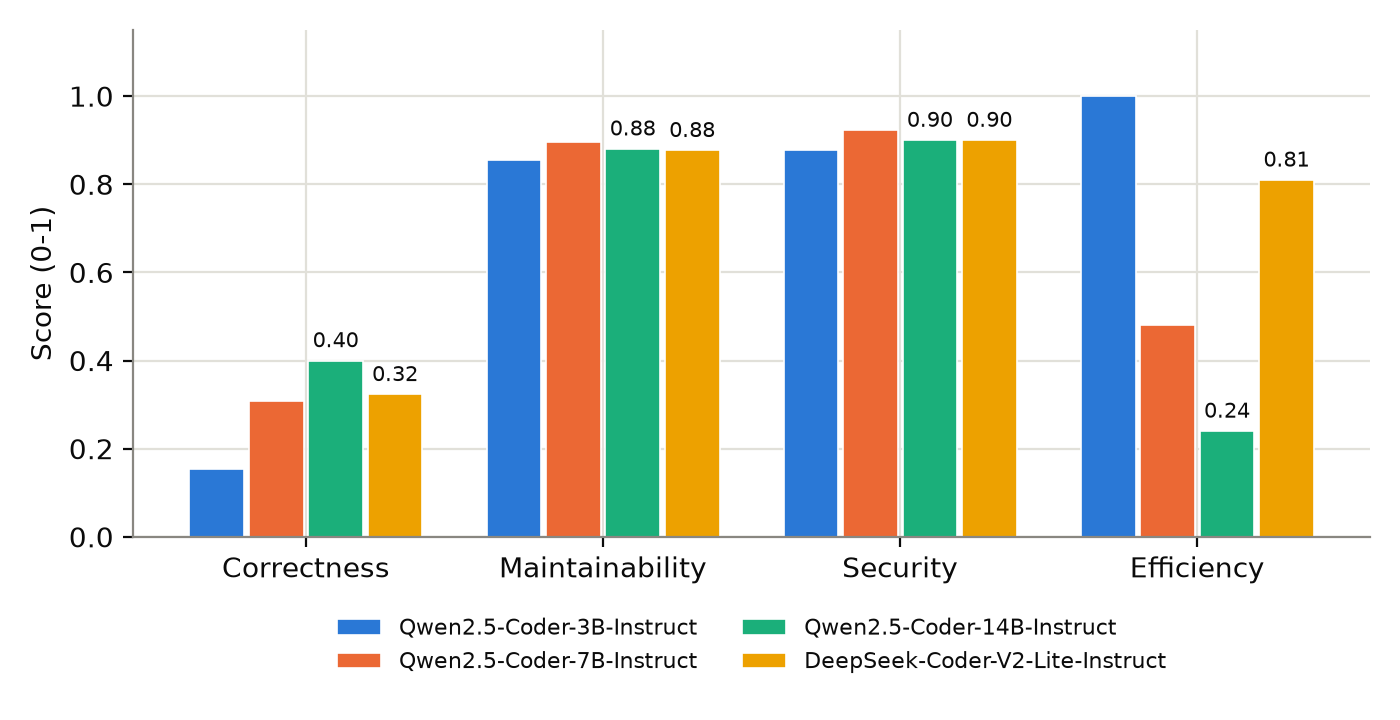}
\caption{Per-model scores across the four quality dimensions.}
\label{fig:dimension_scores}
\end{figure}

Table~III and Fig.~2 report the Quality Index (QI) under the three
weighting schemes. DeepSeek-Coder-V2-Lite ranks first under equal and
correctness-priority weighting, while Qwen2.5-Coder-3B ranks first
under other-priority weighting. These changes show that the overall
ranking depends on how the different quality dimensions are weighted.

\begin{table}[H]
\caption{QUALITY INDEX UNDER DIFFERENT WEIGHTING SCHEMES}
\label{tab:qi}
\centering
\begin{tabular}{lccc}
\hline
Model & Equal & Corr.-Priority & Other-Priority \\
\hline
Qwen2.5-Coder-3B & 0.722 & 0.533 & 0.760 \\
Qwen2.5-Coder-7B & 0.652 & 0.537 & 0.675 \\
Qwen2.5-Coder-14B & 0.605 & 0.537 & 0.619 \\
DeepSeek-Coder-V2-Lite & 0.727 & 0.593 & 0.754 \\
\hline
\end{tabular}
\end{table}
 \begin{figure}[H]
\centering
\includegraphics[width=\columnwidth]{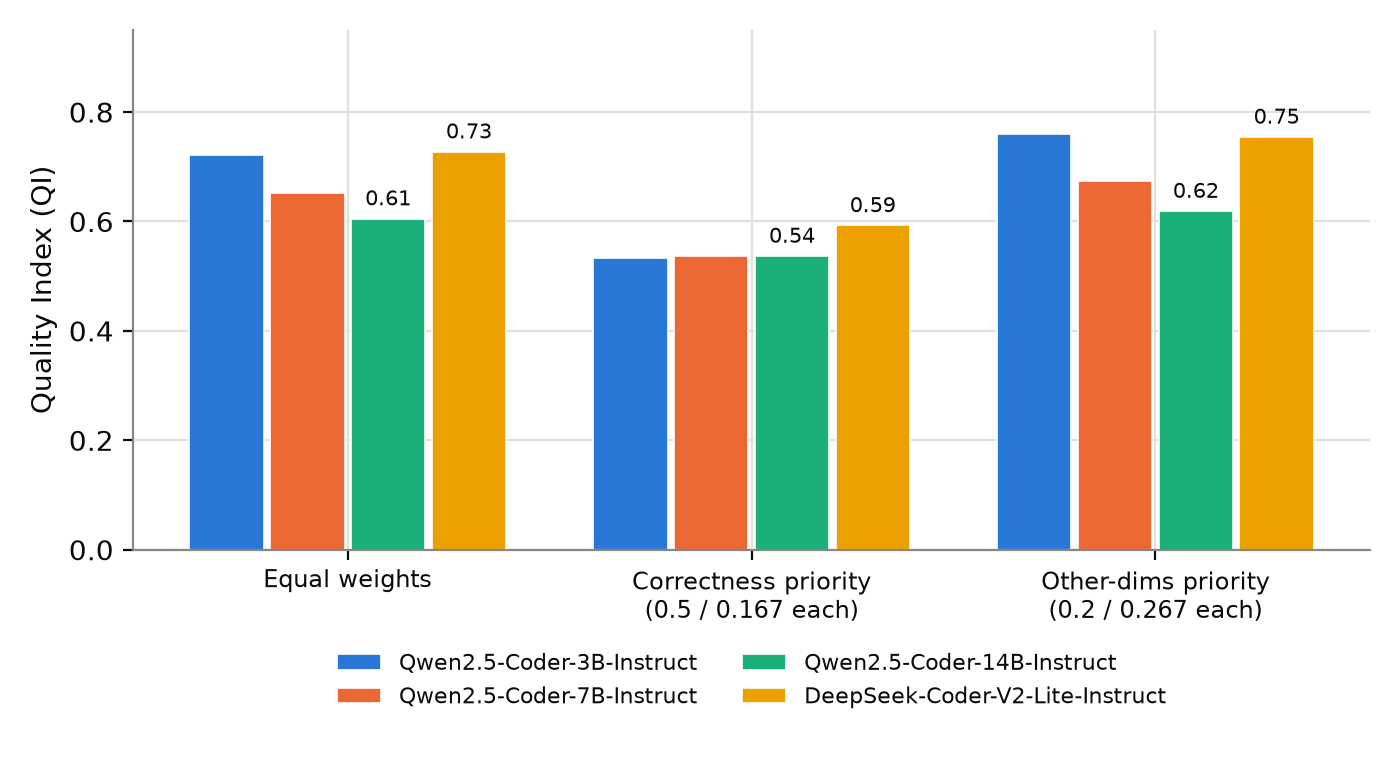}
\caption{Quality Index scores under the three weighting schemes.}
\label{fig:qi}
\end{figure}

Fig.~3 compares generation speed with total and active parameter
counts. Total parameter count is not a consistent predictor of
generation speed: DeepSeek-Coder-V2-Lite, with 16B total parameters,
generates faster than Qwen2.5-Coder-7B despite having more than twice
as many total parameters. Active parameter count provides a more
consistent explanation of the observed generation speeds, particularly
for the sparse MoE model.

\begin{figure}[H]
\centering
\includegraphics[width=\columnwidth]{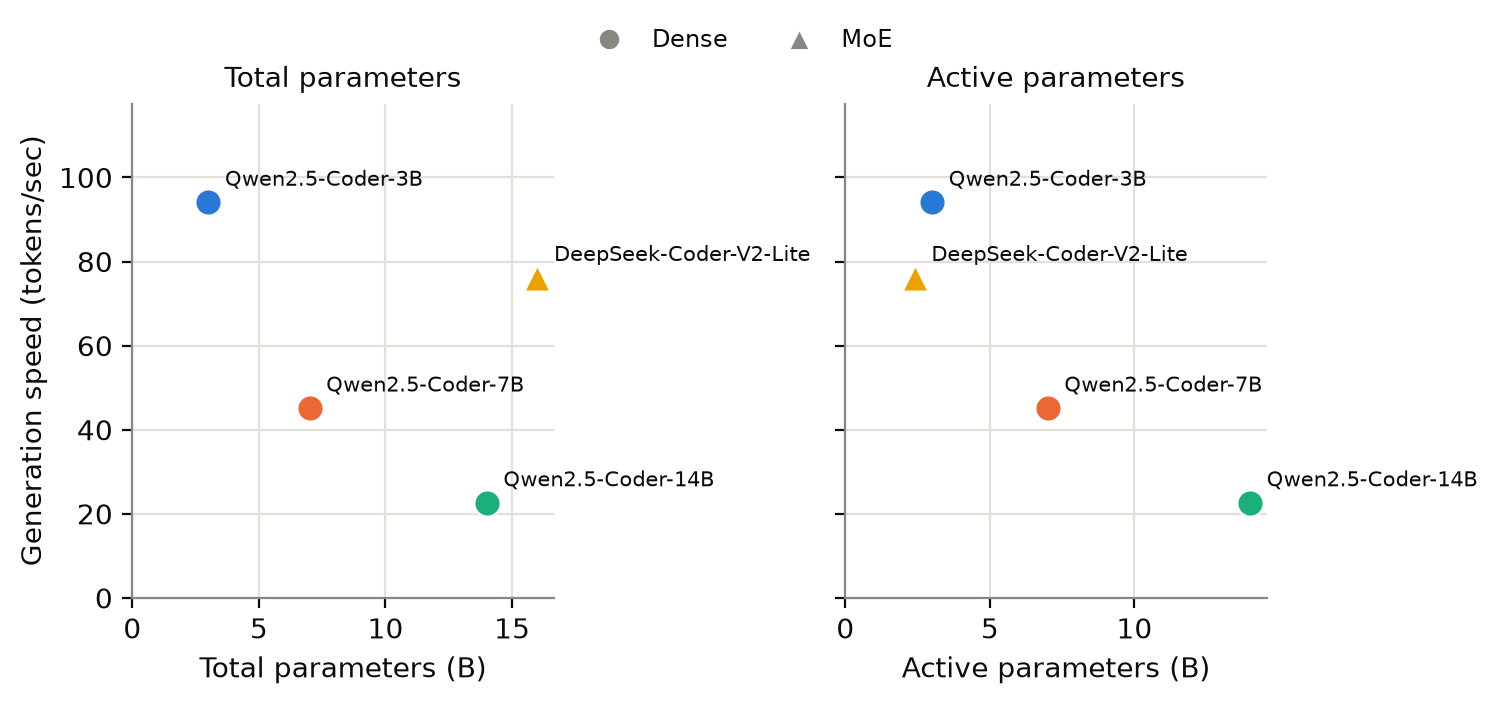}
\caption{Generation speed versus total and active parameter count.}
\label{fig:parameter_efficiency}
\end{figure}

\subsection{Statistical Significance}

Because all four models were evaluated on the same 130 bugs, we compare
their correctness outcomes using McNemar's exact test\cite{mcnemar1947}, which considers
the bugs on which each pair of models disagrees. Table~IV reports the
resulting pairwise comparisons. Since six pairwise
tests are performed, we additionally report Holm-adjusted $p$-values.

Among the three dense models, correctness increases significantly with
model scale: Qwen2.5-Coder-14B outperforms Qwen2.5-Coder-7B, which
outperforms Qwen2.5-Coder-3B, with all pairwise comparisons significant
at $p<0.05$. This is consistent with prior findings that larger models tend to fix more bugs~\cite{xia2023}.

DeepSeek-Coder-V2-Lite is statistically indistinguishable from both
Qwen2.5-Coder-7B ($p=0.83$) and Qwen2.5-Coder-14B ($p=0.053$) despite
using only 2.4B active parameters. Thus, the sparse model achieves
correctness comparable to substantially larger dense models in this
evaluation.

\begin{table}[H]
\caption{PAIRWISE MCNEMAR'S EXACT TEST FOR CORRECTNESS}
\label{tab:mcnemar}
\centering
\begin{tabular}{lcc}
\hline
Comparison & $p$-value & Significant? \\
\hline
DeepSeek vs.\ Qwen2.5-Coder-14B & 0.0525 & No \\
DeepSeek vs.\ Qwen2.5-Coder-7B & 0.8318 & No \\
DeepSeek vs.\ Qwen2.5-Coder-3B & 0.0001 & Yes \\
Qwen2.5-Coder-14B vs.\ 7B & 0.0118 & Yes \\
Qwen2.5-Coder-14B vs.\ 3B & $<0.0001$ & Yes \\
Qwen2.5-Coder-7B vs.\ 3B & $<0.0001$ & Yes \\
\hline
\end{tabular}
\end{table}

Finally, we examined whether Defects4J correctness varies with the size
of the source-level diff. Correctness is 7.3\%, 9.2\%, and 10.5\% for
bugs requiring one, two, and three changed lines, respectively. The
differences are small, suggesting that repair difficulty in this
benchmark is not explained simply by the number of changed source
lines.

\section{Discussion}

\textbf{RQ1.}
The QI framework highlights two aspects that a correctness-only
evaluation would not capture. First, maintainability and security vary
relatively little across the models and do not closely follow
correctness. Thus, higher patch correctness does not necessarily
correspond to higher maintainability or security scores in this
evaluation. Second, the model ranking depends on how the four
dimensions are weighted. DeepSeek-Coder-V2-Lite ranks first under
equal and correctness-priority weighting, while Qwen2.5-Coder-3B ranks
first under other-priority weighting. This shows that the choice of
weights can affect the overall assessment of model quality. We also
observed that min-max normalization can disproportionately affect a
composite score when metrics have different natural scales, as
discussed in Section~III-D.

\textbf{RQ2.}
The MoE model achieves correctness that has no statistically significant difference from the largest dense model evaluated, while having
a higher measured generation efficiency (0.810 vs.\ 0.240). The
statistically significant differences among the three dense models
also show an increase in correctness with parameter scale in this
evaluation. Taken together, these results suggest that the MoE model
can achieve correctness comparable to larger dense models while using
substantially fewer active parameters. However, this should be
interpreted as a comparison within the four models evaluated rather
than as evidence that sparse models generally outperform dense models.

\section{Limitations}

\textbf{Benchmark oracle strength.} QuixBugs' bundled test suites are comparatively small and may not detect subtle incorrectness in a generated patch\cite{ye2018quixbugs}. For Defects4J, correctness is evaluated using trigger tests rather than the full project test suite. Therefore, a patch may pass the evaluated tests while affecting functionality that is not covered by those tests.

\textbf{Sampling.} We report pass@1 using greedy decoding rather than pass@k\cite{chen2021codex} with multiple samples. This choice improves reproducibility and keeps the evaluation tractable, but it may underestimate the performance that could be obtained through repeated sampling.

\textbf{Benchmark coverage.} Ten Defects4J candidates were excluded because their prompts could not be processed reliably within the available context constraints, reducing the selected sample from 100 to 90 bugs. This may limit the representativeness of the Defects4J results, particularly for bugs associated with larger source files.

\textbf{Model coverage.} The evaluation includes three dense models and one MoE model. Additional MoE models at different parameter scales would be needed to determine whether the observed results generalize across sparse architectures.

\section{CONCLUSION}

We presented a multi-dimensional, ISO/IEC 25010-inspired Quality Index
for evaluating APR beyond functional correctness and applied it to three
dense and one sparse code-specialized language model. Model rankings
varied across QI weighting schemes, showing that correctness alone does
not capture all dimensions of patch quality. The evaluated MoE model
achieved correctness statistically indistinguishable from the 7B and 14B
dense models despite using only 2.4B active parameters, while correctness
increased significantly across the three dense model scales. Future work
will evaluate pass@k, additional MoE models, QI confidence intervals, and
direct energy consumption. The code is available at \url{https://github.com/anvishah1/MoE-vs-dense--model-for-APR}

\vspace{12pt}

\end{document}